# Teachers' Perspectives on Integrating AI tools in Classrooms: Insights from the Philippines

**Vanessa B. SIBUG[a], Vicky P. VITAL[a], John Paul P. MIRANDA[a*], Emerson Q. FERNANDO[b], Almer B. GAMBOA[a], Hilene E. HERNANDEZ[c], Joseph Alexander BANSIL[a], Elmer M. PENECILLA[a] & Dina D. GONZALES[a]**
[a]Mexico Campus, Don Honorio Ventura State University, Philippines
[b]Lubao Campus, Don Honorio Ventura State University, Philippines
[c]Bacolor Campus, Don Honorio Ventura State University, Philippines
*jppmiranda@dhvsu.edu.ph

**Abstract:** This study explores the attitudes, reservations, readiness, openness, and general perceptions of Filipino teachers in terms of integrating AI in their classrooms. Results shows that teachers express positive attitude towards integrating AI tools in their classrooms. Despite reporting high level of reservations, teachers believed they are ready and very open in complementing traditional teaching methods with these kinds of technologies. Teachers are very much aware with the potential benefits AI tools can offer to their individual student learning needs. Additionally, teachers in this study reported high level of support from their institutions. Recommendations are offered.

## 1. Introduction

Technology has been a fundamental part of education since the advent of computers (Campbell-Kelly et al., 2023). Artificial intelligence (AI) presents an innovative approach to education that is more personalized, adaptable, inclusive, and engaging (Tapalova & Zhiyenbayeva, 2022; Uunona & Goosen, 2023). AI alongside the vast influx of real-time data, holds significant potential (Luan et al., 2020). AI, a branch of computer science, focuses on developing intelligent machines that mimic human intelligence (Sarker, 2022). The field has expanded rapidly, influencing various aspects of life, including education, where its importance cannot be overstated. AI's applications range from humanoid robots like Yuki and Sophia (Tahiru, 2021) to lip-reading (Deocampo et al., 2024) to game-based learning methods (Hafeez & Hasbi, 2023). However, the integration of new technologies like AI in education introduces new challenges (Luan et al., 2020). Currently, educational research is increasingly oriented towards robotics and intelligent tutoring systems, with a renewed focus on human-centric pedagogies that emphasize cognition and learning (Adel, 2024; Bringula, 2020). AI techniques have facilitated the creation of smart learning environments that employ deep learning, recommendation systems, analytics, and classification to meet educational objectives (Bringula, 2023; Zhai et al., 2021). According to Holmes and Tuomi (2022), tools like Google Docs and Sheets represent the forefront of AI-assisted educational technologies (Malhan et al., 2024). Additionally, platforms such as YouTube, TikTok, WeChat, and WhatsApp are increasingly used to enhance learning (Holmes & Tuomi, 2022), although the efficacy of AI technologies like activity trackers remains under debate (Nguyen et al., 2023; Zhai et al., 2021).

In classroom settings, AI tools have markedly improved the educational experience (Gamis, 2024). They facilitate complex activities, provide immediate feedback, and reduce the administrative burden on teachers and further enhance both educational and administrative processes (Lawrence et al., 2024). AI also enriches learning by analyzing cognitive behaviors to support personalized instruction and by improving students' thinking skills and reading abilities (Alharbi, 2023). Furthermore, AI has the potential to improve education through optimization of enrollment processes, resource management including the enhancement of online training programs (Lukianets & Lukianets, 2023). Furthermore, understanding teachers' perceptions of integrating AI tools into traditional classrooms is crucial for their successful implementation and adoption (Choi et al., 2023). Teachers plays important roles in pioneering educational innovations and shaping new pathways. Their attitudes towards AI suggestively influence the effectiveness and acceptance of its integration into educational processes, which is vital for creating enriching learning environments and enhancing student experiences (Alqahtani et al., 2021; Polak et al., 2022). Past researches have highlighted that the success of integrating technology in education largely depends on teachers' beliefs and attitudes towards such tools (Dogan et al., 2021). Their perceptions are essential in determining the effectiveness of integrating information and communication technology in classrooms (Abel et al., 2022). Furthermore, a teacher's readiness to embrace new technologies is a decisive factor in their effective integration into classroom instruction (Kim & Kwon, 2023). Based on these, this study aims to explore the perspectives of Filipino teachers on integrating AI tools like ChatGPT, Grammarly, Canva Magic Classroom, Google Gemini etc. in traditional classrooms, focusing on their attitudes, reservations, readiness, and enabling factors for the successful adoption of these technologies in education.

## 2. Method

This study employed a descriptive research design. The survey was conducted from November 2023 to March 2024 with 292 teachers participating. The survey instrument was validated by two professors specializing in educational technology, both of whom hold doctoral degrees in Educational Management and have at least five years of teaching experience. Additionally, the instrument underwent reliability testing involving 32 participants in a pilot study, who were subsequently excluded from the main survey. The results of the analysis confirmed the reliability of the instrument, with a Cronbach's alpha of 0.943. Individual Cronbach's alpha scores for each construct were reported as follows: Attitudes ($\alpha = 0.859$), Reservations ($\alpha = 0.905$), Readiness ($\alpha = 0.958$), Openness ($\alpha = 0.924$), General Perceptions ($\alpha = 0.962$), and Enabling Factors ($\alpha = 0.942$). The respondents of the study were age ranged from 27 to 47 years old. On average, the teachers in this study have three to 19 years of teaching experience. In terms of sex distribution, majority of the respondents were female (74.7%) and work in public schools (96.2%). Most of them were married (57.5%) followed by single (38.7%). Most of them are licensed professionals (90.1%). For academic rank, most of them are ranked as teachers (79.1%) where almost half of them are primarily teaching in junior high school (48.3%) followed by elementary (17.8%), senior high school (17.8%), and college (16.1%). More than half of them holds a bachelor's degree (62%) and only a fraction of them holds doctorate degree (3.4%). For technology access, a high number of the respondents have access to laptops (86.6%) and smartphones (79.1%), with fewer having desktops (40.1%) and tablets (17.1%) and most of them learn about professional trends through conferences, seminars, or workshops (83.2%), followed by social media (72.9%) and webinars (58.6%). Journals and magazines are less commonly used (28.1%).

## 3. Results and Discussion

### *3.1 Teachers' attitudes towards AI*

Teachers believed integrating AI can enhance student learning (Mean = 4.97, SD = 0.941). This suggests that teachers view AI tools as capable of personalizing learning by providing adaptive resources and feedback tailored to the individual pacing and learning styles of students. Regarding their confidence in using AI tools in educational settings, teachers generally feel capable (Mean = 4.87, SD = 0.901). Additionally, they recognize the potential of

AI to add value to traditional teaching methods and improve learning outcomes (Mean = 5.03, SD = 0.883). This indicates a belief that AI will complement rather than replace traditional teaching methods. Furthermore, the high level of excitement about the dynamic learning possibilities introduced by AI is evident (Mean = 5.03, SD = 0.892), reflecting a positive attitude toward the innovative capabilities of AI in education. This enthusiasm suggests that teachers are keen on adopting AI-driven innovations in teaching strategies and integrating new technologies into the classroom. It also points to a potential shift in educational paradigms towards more student-centered and technology-driven approaches. Furthermore, teachers strongly believed that AI can effectively meet the unique learning needs of individual students (Mean = 4.97, SD = 0.902). This shows that teachers recognize the potential of AI to support differentiated instruction, which could lead to more inclusive educational practices that accommodate diverse learning needs.

*3.2 Reservations about AI integration*

Teachers express concerns about potential technical issues that could arise from using AI tools in the classroom (Mean = 4.90, SD = 0.916). These issues include software malfunctions, hardware compatibility problems, and internet connectivity issues. This suggests that school administrations may need to enhance the technology support provided to teachers, including upgrading the technology infrastructure within their institutions. Adequate technical support could alleviate these concerns and encourage more teachers to embrace AI technologies in their teaching. Additionally, teachers are worried about an over-reliance on AI tools potentially weakening their personal connections with students (Mean = 0.75, SD = 1.073). This concern highlights the belief that the teacher-student relationship is crucial for effective learning. A balanced integration of AI tools should ensure that they complement rather than replace personal interactions. For privacy and security risks, teachers are apprehensive about using AI in education, particularly concerning the handling of sensitive student data (Mean = 4.86, SD = 0.856). This calls for strict data protection measures within institutions, implementing appropriate policies to safeguard both teachers' and students' information. Establishing clear protocols can help mitigate these concerns and aid in the successful implementation of AI tools in the classroom. Moreover, there is concern about the learning curve associated with using AI tools (Mean = 4.91, SD = 0.743). Teachers anticipate challenges in effectively using AI tools, likely due to its disruptive nature and the novelty of the technology. This high level of concern highlights the need for professional development tailored to using AI in the classroom. In addition to these, teachers are also apprehensive about potential resistance to AI tools from students or parents (Mean = 4.89, SD = 0.869). This may stem from fears about the impersonalization of teaching or concerns about student data privacy. Addressing these issues requires effective communication and community engagement during the rollout of any AI tools. Educating all stakeholders could reduce these reservations and potentially foster a supportive environment for technological advancements in the classroom.

*3.3 Readiness for AI implementation*

Teachers reported a have high confidence in their ability to use AI tools effectively in their teaching (Mean = 4.60, SD = 0.893), indicating a strong belief in their personal competence with these technologies. This confidence is crucial for the successful implementation and adoption of new technologies in the classroom. However, while teachers believe they are somewhat well-trained to integrate AI tools, the higher standard deviation (Mean = 4.34, SD = 1.031) suggests varying levels of readiness among them. Addressing this issue may require more consistent and tailored training programs that enhance teachers' preparedness and comfort with using and integrating AI. Similarly, while teachers believed they have the necessary resources and support to implement AI in their classrooms (Mean = 4.34, SD = 1.018), this high variability as indicated by the standard deviation suggests that not all teachers feel equally supported. Schools may need to ensure that all teachers have access to the necessary resources and support systems to facilitate the seamless integration of AI tools in the classroom. Additionally, there is a general confidence in handling technical problems associated with AI tools (Mean = 4.22, SD = 1.052). By providing adequate training relevant to these issues could increase teachers' readiness and confidence. Teachers are generally comfortable with modifying their teaching methods to incorporate AI tools (Mean = 4.58, SD = 0.864). This suggests that teachers are adaptive and very open to exploring new pedagogical approaches that align with the use of AI tools. Furthermore, teachers clearly understand how AI can enhance their current teaching methods (Mean = 4.60, SD = 0.949). This result indicates that teachers have a solid grasp of the potential benefits of AI in the classroom.

### 3.4 Openness to AI adoption

Teachers indicated that they were very open to experimenting with AI tools to enhance student learning, demonstrating a proactive attitude towards adopting technological innovations such as AI (Mean = 5.10, SD = 0.832). This high level of openness suggests that the introduction of AI tools in the classroom could be met with a positive reception and active engagement from teachers. Furthermore, there was a strong eagerness among teachers to discover and understand new AI tools (Mean = 5.06, SD = 0.787). This eagerness likely indicates a commitment to professional development and continuous learning. Educational administrators should have consider investing more in professional development and training focused on AI integration, responding to this demonstrated interest. Moreover, teachers were also willing to adapt their teaching styles to incorporate AI tools (Mean = 5.06, SD = 0.818). This suggests that they possessed the flexibility and forward-thinking approach necessary for integrating new technologies into their teaching methods. It was likely that teachers believed AI tools could facilitate dynamic and effective learning environments. Moreover, teachers strongly believed in the transformative potential of AI in education (Mean = 4.95, SD = 0.825). Their excitement and enthusiasm for participating in this change could have been a driving force for embracing AI-based educational integration within their classrooms. Additionally, teachers were receptive to feedback and suggestions on how to better use AI in their classrooms (Mean = 5.10, SD = 0.811). This indicates an acknowledgement of the adaptive and collaborative needs for successful implementation and usage of such technologies. Such openness to feedback could have further enhance the effectiveness of AI tool implementation and possibly promoting a culture of continuous improvement and collaborative problem-solving in educational settings.

### 3.5 General perceptions about AI tools

Teachers believed that AI tools can tailor learning experiences to the individual needs of students in traditional classrooms (Mean = 4.91, SD = 0.842). This strong consensus highlights the increasing importance of AI in education. Additionally, teachers view AI not as a replacement but as a beneficial complement to traditional teaching methods (Mean = 4.88, SD = 0.841) - a similar result indicated above. This recognition of AI as a supportive technology suggests its potential integration into existing educational frameworks, enhancing teaching practices without displacing traditional methods. In terms of boosting student engagement and motivation (Mean = 4.86, SD = 0.844), teachers believed that AI can positively enhance these aspects. They suggest that AI tools can make learning more engaging, dynamic, enjoyable, and interactive. Furthermore, teachers consider AI a helpful tool in addressing the diverse learning needs of their classrooms (Mean = 4.88, SD = 0.805). The capability of AI tools to cater to various learning styles could lead to more inclusive educational practices. Teachers also believed that AI tools can streamline administrative and instructional tasks. They feel that the time saved by automating these tasks can be redirected to more personal interactions with students, thereby improving the quality of education (Mean = 4.87, SD = 0.813). These kinds of perceptions encouraged teachers to adopt AI tools to automate routine tasks, allowing them to focus more on pedagogical aspects within their classrooms.

### 3.6 Enabling factors for successful AI integration

Regarding social support, teachers feel well-supported by their schools in using AI tools (Mean = 4.43, SD = 1.041). This suggests that such schools are likely encouraging this integration.

Continual advocacy and support from school leaders could further enhance teacher engagement with AI technologies. Similarly, teachers perceive strong support from school administrators in providing the necessary resources for AI implementation in their classrooms (Mean = 4.42, SD = 1.051), despite high variability in social support. For policy support, teachers are relatively well-informed about their school's policies for ethical and responsible use of AI (Mean = 4.44, SD = 1.008). They also show a strong commitment to following ethical guidelines for integrating (Mean = 4.47, SD = 0.979). This indicates that ongoing education on AI ethics and policy can help maintain high standards for its responsible use. Strengthening existing policy frameworks could further support teachers by providing clear guidelines and expectations. In terms of infrastructure, concerns arise regarding the adequacy of internet infrastructure (Mean = 3.95, SD = 1.200) and the suitability of existing technology infrastructure for optimal AI use (Mean = 3.97, SD = 1.145). These concerns suggest that investment in upgrading internet capabilities and technological facilities is necessary to support the sophisticated demands of AI technologies. Regarding personal capacity, teachers generally have access to the necessary devices for AI use (Mean = 4.61, SD = 0.980) and feel competent in their technical skills (Mean = 4.46, SD = 0.882). However, access to necessary software and applications shows some high variability (Mean = 4.29, SD = 1.030). Maintaining or increasing access to updated devices and providing ongoing technical trainings can enhance effective AI usage and continued integration. Streamlining access to relevant technologies and applications will ensure that teachers will continue to integrate AI effectively into their teaching practices.

## 4. Conclusion, Recommendations, and Limitations

Teachers are highly positive about incorporating AI tools into their classrooms. They recognized the potential of the tool to personalize and enhance learning. They view these tools as valuable additions to traditional methods, despite concerns about reduced personal interaction and privacy. Teachers are eager to use AI to streamline administrative tasks which allows them to have more time for direct student engagement. The study highlighted the need for robust technical support, data protection, and professional development to address the learning curve and resistance. It emphasized the necessity of balanced AI use that prioritizes personal interactions. Additionally, this study also shows the need for consistent, tailored training programs, and investments in internet and technology infrastructure to ensure all educators are equally prepared. Continuous updates and training are recommended to address disparities in access to essential software. However, limitation for this study include the potential bias since the majority of respondents were from public schools. The descriptive nature of this study, the lack of depth in exploring causality, and the geographic focus on the Philippines may limits the generalizability of the results.